\documentclass[conference]{IEEEtran}
\usepackage{graphicx}
\usepackage[hidelinks]{hyperref}
\usepackage{amsmath}
\usepackage{amssymb}
\usepackage[]{algorithm2e}
\usepackage{subfigure}
\makeatletter
\newcommand*{\rom}[1]{\expandafter\@slowromancap\romannumeral #1@}
\makeatother
\usepackage{enumerate}

\usepackage{lipsum}

\makeatletter
\newcommand{\nosemic}{\renewcommand{\@endalgocfline}{\relax}}
\newcommand{\dosemic}{\renewcommand{\@endalgocfline}{\algocf@endline}}
\makeatother

\usepackage{cite}

\ifCLASSINFOpdf
\else
\fi
\begin{document}
%
\title{AirScript - Creating Documents in Air}



 
\author{\IEEEauthorblockN{Ayushman Dash\IEEEauthorrefmark{1}\IEEEauthorrefmark{3},
Amit Sahu\IEEEauthorrefmark{1}\IEEEauthorrefmark{3},
Rajveer Shringi\IEEEauthorrefmark{1}\IEEEauthorrefmark{3},
John Gamboa\IEEEauthorrefmark{4}\\
Muhammad Zeshan Afzal\IEEEauthorrefmark{4},
Muhammad Imran Malik\IEEEauthorrefmark{2}, Sheraz Ahmed\IEEEauthorrefmark{2} and
Andreas Dengel\IEEEauthorrefmark{2}}
\IEEEauthorblockA{\IEEEauthorrefmark{3}Knowledge-Based Systems Group,\\
Department of Computer Science, University of Kaiserslautern,\\
P.O. Box 3049, 67653 Kaiserslautern, Germany}
\IEEEauthorblockA{\IEEEauthorrefmark{4}MindGarage, University of Kaiserslautern, Germany}
\IEEEauthorblockA{\IEEEauthorrefmark{2}German Research Center for AI (DFKI)\\
Knowledge Management Department,\\
Kaiserslautern, Germany}}



\maketitle


\begin{abstract}

This paper presents a novel approach, called AirScript, for creating, recognizing and visualizing documents in air. We present a novel algorithm, called 2-DifViz, that converts the hand movements in air (captured by a Myo-armband worn by a user) into a sequence of $x, y$ coordinates on a 2D Cartesian plane, and visualizes them on a canvas. 
Existing sensor-based approaches either do not provide visual feedback or represent the recognized characters using prefixed templates.
In contrast, AirScript stands out by giving freedom of movement to the user, as well as by providing a real-time visual feedback of the written characters, making the interaction natural.
AirScript provides a recognition module to predict the content of the document created in air. To do so, we present a novel approach based on deep learning, which uses the sensor data and the visualizations created by 2-DifViz. The recognition module consists of a Convolutional Neural Network (CNN) 
and two Gated Recurrent Unit (GRU) Networks.
The output from these three networks is fused to get the final prediction about the characters written in air. 
AirScript can be used in highly sophisticated environments like a smart classroom, a smart factory or a smart laboratory, where it would enable people to annotate pieces of texts wherever they want without any reference surface. We have evaluated AirScript against various well-known learning models (HMM, KNN, SVM, etc.) on the data of 12 participants. Evaluation results show that the recognition module of AirScript largely outperforms all of these models by achieving an accuracy of 91.7\% in a person independent evaluation and a 96.7\% accuracy in a person dependent evaluation.

\end{abstract}


%
\IEEEpeerreviewmaketitle

\section{Introduction}
{\let\thefootnote\relax\footnote{$^*$These authors contributed equally to this work.}}






In the last few decades, the definition of document
has been the topic of significant discussion.
Some authors have restricted documents to ``things that we can
read" \cite{macdonald2001using}, while others have extended it to
anything that \textit{functions} as a source of evidence
\cite{buckland1998digital}.
We consider a document as any resource for
furnishing information evidence or proving the information
authenticity\footnote{Keynote speech by Andreas Dengel (DAS 2014): \url{https://das2014.sciencesconf.org/resource/page/id/5}}. However, the value of documents ``cannot be fully estimated by just looking at their contents''\footnote{Keynote speech by Koichi Kise (ICDAR 2015): \url{http://www.iapr.org/archives/icdar2015/index.html\%3Fp=372.html}} as the activities related to the documents provide key information related to them.
These discussions are important because they have implications on
the strategies that can be used to generate documents in air.

During the Information Age, the media where documents are created has undergone a fast transition from traditional paper-based methods to any digital device.
Documents are nowadays created in laptops, PCs and smartphones, by means of text editors and drawing tools, or alternatively generated in real-time on flat surfaces able to perform handwriting recognition. 
However, despite the progress, all of these methods are limited in that they restrict the region where the input is received to a given surface of reference.

\begin{figure}[h]
\centering
\includegraphics[width=.5\textwidth]{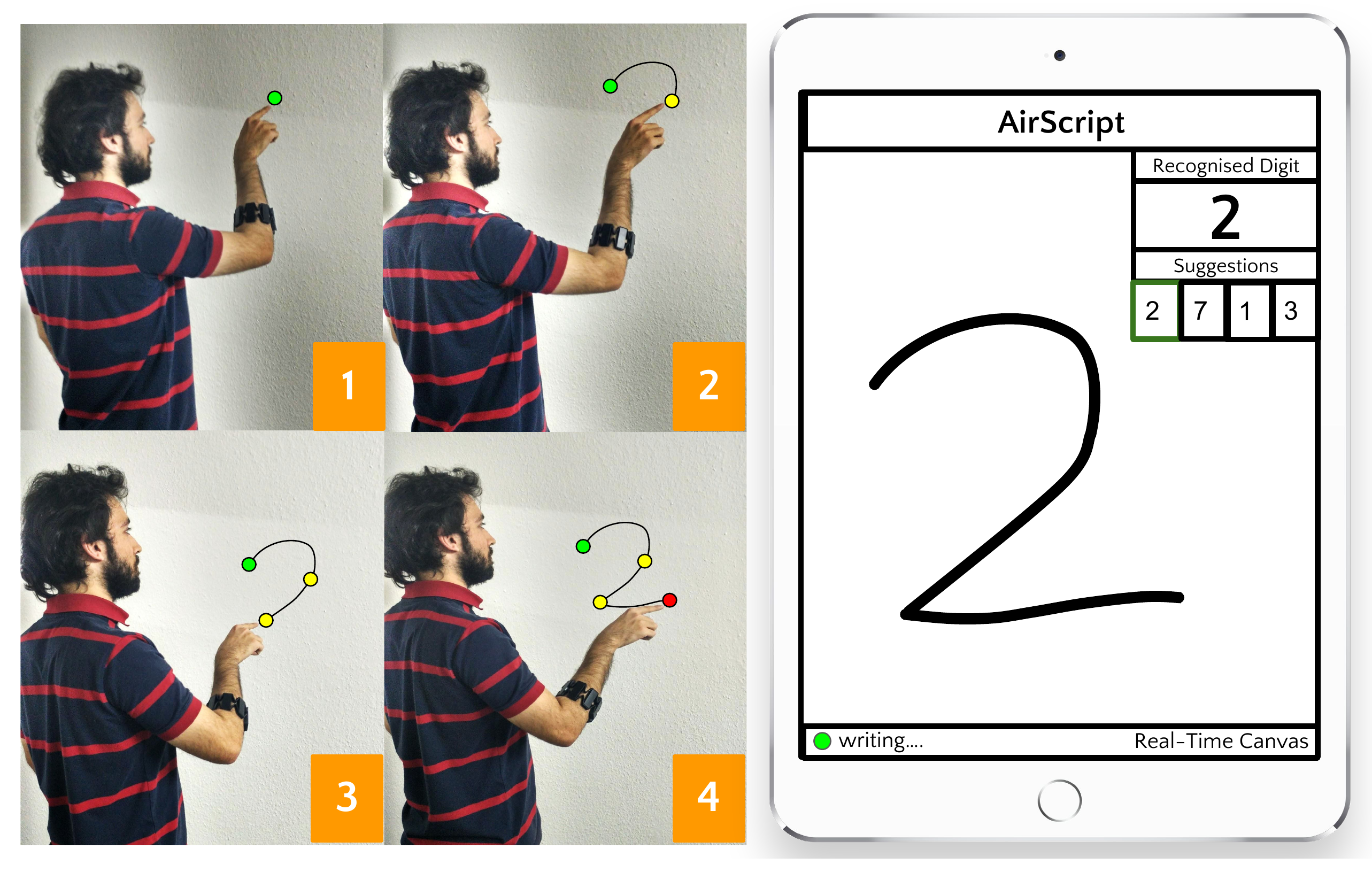}
\caption{A smart classroom scenario that we envision, where AirScript can be used for writing in air by just wearing a Myo-armband. The images on the left show how a person writes in air (in our case, digits). The numeric label in orange represents the sequence of hand movements. While the person writes in air, the Myo-armband captures raw IMU and EMG signals from the arm and sends it to AirScript, running on a digital device. AirScript gives a realistic visual feedback to the user in real time on that device, showing what the user wrote in air. It also recognizes the written digit, giving possible suggestions for it.}
\label{fig:exampleScene}
\end{figure}

In this article we introduce AirScript, a novel approach for document creation in air, whereby a sensor device is attached to the user's arm, capturing its movements. This way, we eliminate the dependency on a reference surface, overcoming a major drawback of the previous methods.
Our method has the potential to enable people to annotate pieces of texts wherever they want, with complete freedom of movement. It could be used in highly sophisticated environments, such as a smart classroom or a smart factory. The recognized content could then be easily displayed on a board or any other canvas if the user so likes.

AirScript is composed of two modules. Its visualization module, \textbf{2-DifViz}, projects the multi-dimensional sensor data onto a 2D surface, producing a realistic visualization of the input. This visualization is then used, along with the raw sensor data, by AirScript's \textbf{recognition module} to predict the content of the hand movement.
We developed our proof of concept using a Myo-armband, as described in Section~\ref{sec:myoab}.
Finally, our recognition module performs Handwritten Digits Recognition in Air (HDRA). The data sensed by the Myo-armband is fed into a Gated Recurrent Unit (GRU) network. 2-DifViz produces differential features that are fed to another GRU Network, as well as used to produce visualizations that are fed to a Convolutional Neural Network (CNN). The result of the three networks is then fused to produce a prediction of the digit written in air. It was found that this fusion made the classification results more robust when testing the model with several participants. Figure~\ref{fig:exampleScene} shows an example scenario, in which we envision that our approach could be employed.
A demo video of the = is also available.\footnote{\url{https://drive.google.com/file/d/0B5xrMEupo2dPbEk4ekNVOUhMOWs}}


\begin{figure}[t]
\centering
\includegraphics[width=8cm]{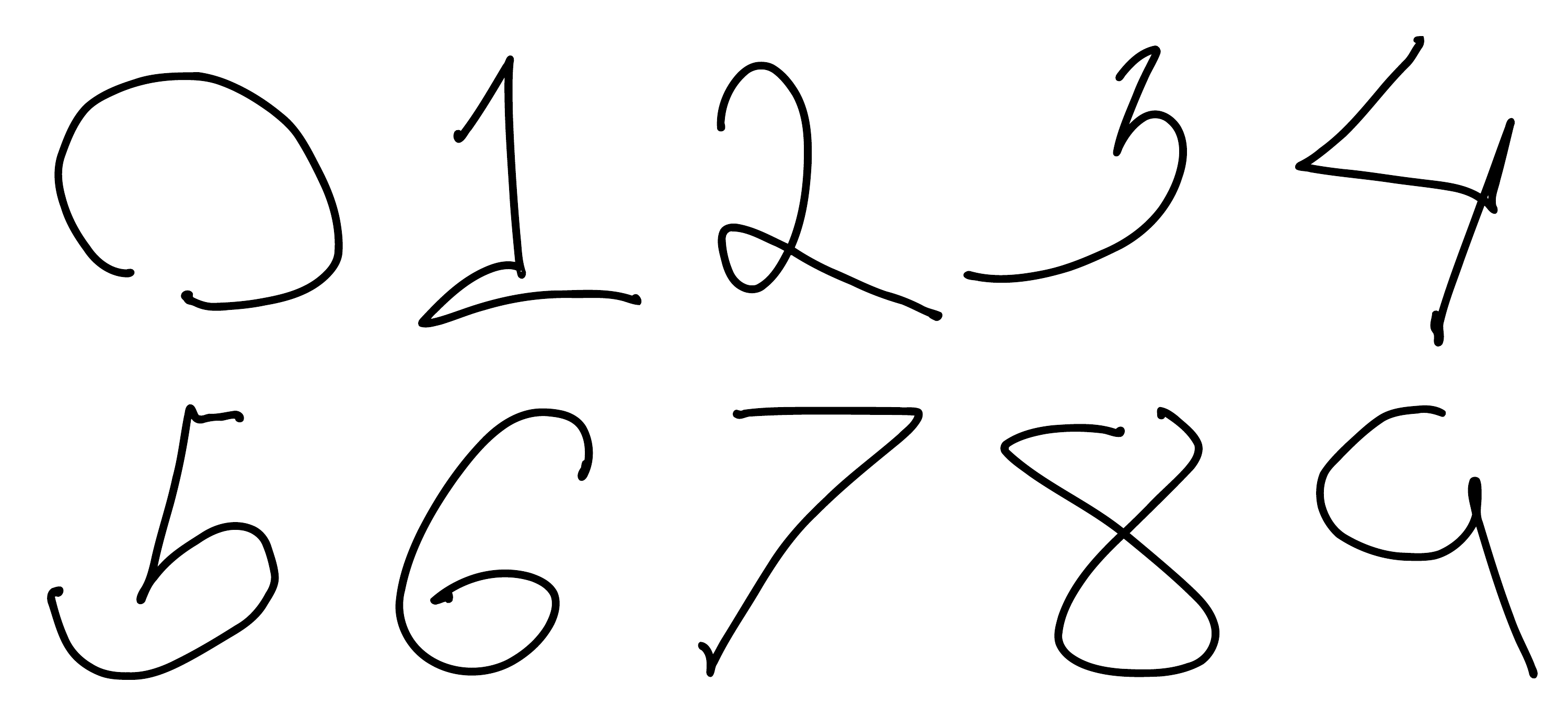}
\caption{Visualization of the hand movements for the digits that were drawn in air. Section \ref{method3} describes how the hand movements were converted into a sequence of $x, y$ coordinates, represented in the form of images. }
\label{fig:vizHandDigits}
\end{figure}

To gauge the performance of AirScript, it is evaluated on HDRA data of 12 different participants.  
Evaluation results reveal that  AirScript achieves an average accuracy of 96.7\% on a person dependent evaluation, and 91.7\% on a person independent evaluation. Furthermore, the visualization of hand movements produced by AirScript are also very realistic, which makes it suitable for using in real scenarios. 
Figure~\ref{fig:vizHandDigits} shows the visualization of the hand movements of the participants while they were drawing the digits in air. This not just shows the potential of the proposed HDRA model but also makes it evident that the recognized digits can also be converted into a reproducible human readable format.



\section{Related Work}
\label{relatedWork}
A lot of work has been done on extending and simplifying the process of creating documents. Examples include using gesture-based input control\cite{gestureInputShadow,brainyHand,wearableGesture,kinect}, swipe-based input methods\cite{swipeKeypad}, voice-based input methods\cite{deepSpeech} and even interaction methods on imaginary surfaces\cite{imaginarySurface}.
Specially relevant to this work are methods that create a Virtual Reality\footnote{For example, \url{https://www.tiltbrush.com/}} environment where the user explicitly inputs information. Because of their Virtual Reality nature, these methods are inadequate in environments where the user has to interact with real world objects, such as a smart factory or a smart class. Our approach differs from these methods in that the user remains in the real world.

Handwriting Recognition in Air (HWRA) has been performed by
Amma et al. \cite{airWriting}, using a prototype glove with an embedded IMU sensor, showing promising results.
They combine a Hidden Markov Model (HMM) with a language model, and achieve a word error rate of 11\% on a person independent evaluation and 3\% on a person dependent evaluation. However, their method does not give any visual or haptic feedback to the user. 
Alternative computer vision-based approaches relying on finger tracking\footnote{For example, \url{https://www.leapmotion.com/}} or multi-camera 3D hand tracking have been used for HWRA\cite{visionBasedMidAir, handwritingOnAirLeap1,onTheFly} but face problems similar to those of the finger tracking approaches.
These methods are dependent on a tracking device that has to be on the line of sight, restricting the freedom of movement of the user. To our knowledge, Deep Learning methods have not been explored for HWRA.

In opposition to HWRA, handwriting recognition on surface using Deep Learning models like Bidirectional LSTMs\cite{bidirLstm}, Connectionist Temporal Classifiers\cite{gravesCtc}, and  Multidimentional RNNs\cite{multiRnn} have outperformed other baseline models (\textit{e.g.}, \cite{liwicki2006hmm}). Similar Deep Learning models, such as Convolutional Neural Networks\cite{convLstm} and LSTMs\cite{lstmGesture}, have also shown improved results in the domain of gesture recognition.


\section{Data Acquisition}
\label{dataCollection}

To train and test the performance of AirScript, a dataset was collected from 12 right handed participants while they drew digits in air. The recording includes raw IMU and EMG signals from a Myo-armband as well as their ground truth labels.
For data acquisition and visualization, a complete  Graphical User Interface based solution called Pewter\footnote{\href{https://github.com/sigvoiced/pewter}{https://github.com/sigvoiced/pewter}} was developed. 


The Myo-armband has an Inertial Measurement Unit (IMU) that consists of a 3-dimensional accelerometer that measures the non-gravity acceleration, a 3-dimensional gyroscope that measures the angular momentum, and a magnetometer that measures orientation w.r.t. the Earth's magnetic field. A 10-dimensional vector $\mathcal{M}$ is acquired from the IMU at a sampling rate of $50Hz$.
Each of these sensors acts like a function $f:~T\rightarrow\mathbb{R}$ that maps a timestep to a real value. For a duration of time, they form a sequence (represented as vector) of real values corresponding to the digit written  in that time.

The participants were asked to wear the device on the right arm in accordance with the Myo-armband instructions\footnote{https://s3.amazonaws.com/thalmicdownloads/information+guide/important-information-guide-v03.pdf}. 
To avoid fatigue and priming, the data for each digit was collected in three phases. In Phase-\rom{1} and Phase-\rom{2}, 3 iterations of every digit were conducted. Phase-\rom{3} consisted of 4 iterations.

Due to unusual vibrations in the Myo-armband during the data collection process, 30 data samples were visualized using Pewter and removed manually from the dataset.
Hence, the final dataset contained 1270 samples in total.

\begin{figure*}[t]
\includegraphics[width=18cm]{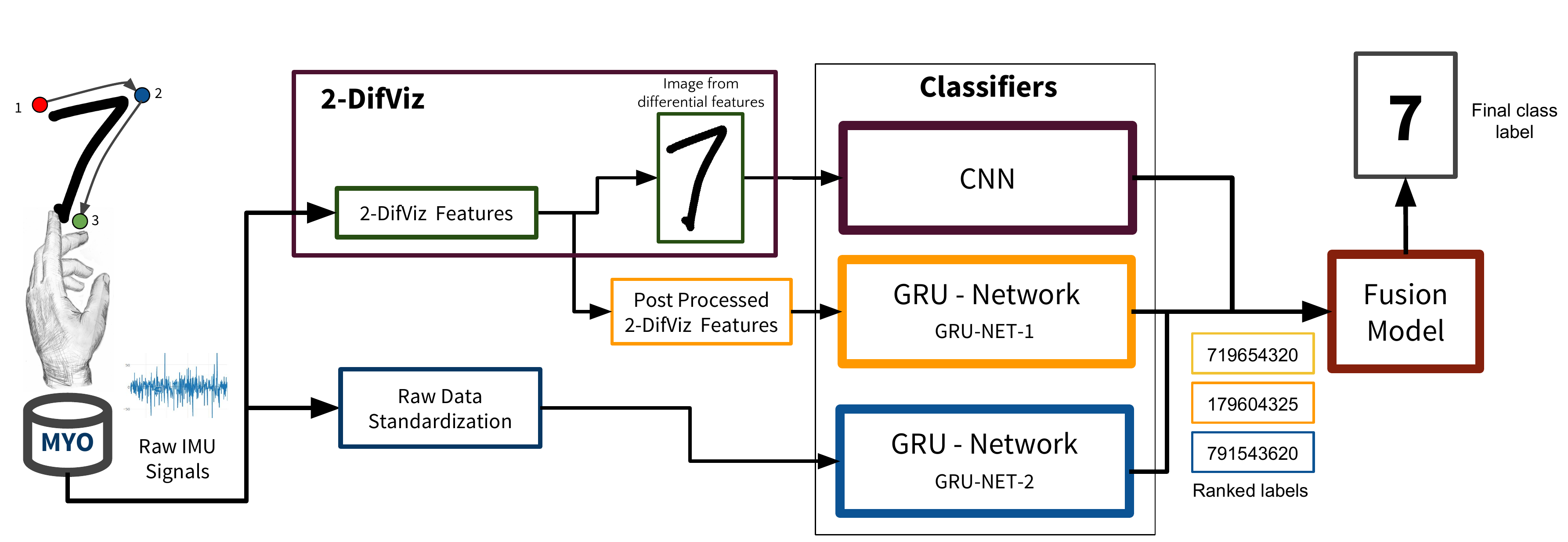}
\caption{The architecture of the proposed HDRA fusion model. The colored dots on the top-left show the sequence of hand movements. The red dot represents the starting point, the blue dot represents an intermediate point and the green dot represents the end of the hand movement. A sequence of $x,y$ coordinates representing the user's hand movements is extracted from the raw IMU signals using 2-DifViz. These features are projected onto a 2D plane, forming the digit the user drew, and fed to a Convolutional Neural Network (CNN). They are also post-processed and fed to a Gated Recurrent Unit (GRU) Network. Additionally, the raw signals are standardized and fed to a separate GRU Network. Finally, the output of the three classifiers is fused to produce the final prediction. 2-DifViz is the Visualization module of AirScript (Phase-\rom{1}), and the CNN, GRU-NET-1 and GRU-NET-2 compose its recognition module (Phase-\rom{2}).
}
\label{fig:recognitionArch}
\end{figure*}

\section{AirScript: The Presented Approach}
\label{methodology}
To generate documents in air we propose a two phase process that breaks down the problem into two different tasks:

\begin{enumerate}
\item \textbf{Phase-\rom{1} (2-DifViz)}: the hand movements are converted into a realistic visualization of the digit written in air.

\item \textbf{Phase-\rom{2} (HDRA)}: handwritten digits in air are recognized using a fused classifier.  
\end{enumerate}

Figure~\ref{fig:recognitionArch} shows the complete workflow of AirScript, in which the raw IMU data from the Myo-armband is processed using 2-DifViz and a signal standardization pipeline. The processed data is then fed to a recognition module consisting of three classifiers. Each of these classifiers provides a list of ranked results. These are then fused and a final prediction is generated. Because the mistakes committed by the three classifiers are generally different, this fusion was found to improve the robustness of the model.




\subsection{Myo-Armband} \label{sec:myoab}

The Myo-armband is an unobtrusive sensor device easily available and integrable to several platforms through its off-the-shelf SDKs. Its Inertial Measurement Unit (IMU) senses the orientation, acceleration and angular velocity of the arm at any given moment. Additionally, the arm's muscle activity is captured by 8 Electromyography (EMG) pods embedded in the device.


Let $\mathcal{D}=\{d_i\mid i=1,...,n\}$ represent our HDRA dataset, where $n$ is the number of data instances in $\mathcal{D}$ (in our case 1270). Each data instance $d_i=(\mathcal{M}_i,\mathcal{E}_i, L_i)$ is a tuple consisting of a time-series $\mathcal{M}_i$ representing the IMU sensor data, a time-series $\mathcal{E}_i$ representing the EMG sensor data, and a class label $L_i \in \{0,...,9\}$. For our models we use only the IMU sensor data.
Every $\mathcal{M}_i=\{\mathcal{M}_i^{(1)}, \mathcal{M}_i^{(2)},...,\mathcal{M}_i^{(\tau_{i})}\}$ is a time-series consisting of $\tau_i$ time-steps (\textit{i.e.}, $|\mathcal{M}_i|=\tau_i$), and each element $\mathcal{M}_i^{(t)} = \bigl[ \begin{matrix}
                        \textbf{a}_i^t&
                        \textbf{g}_i^t&
                        \textbf{q}_i^t
                        \end{matrix} \bigr]$. The vectors $\textbf{a}_i^t \in \mathbb{R}^3$ and $\textbf{g}_i^t \in \mathbb{R}^3$ are the 3 axes of the accelerometer and the gyroscope, respectively. Similarly, $\textbf{q}_i^t \in \mathbb{R}^4$ denotes a quaternion representing the orientation. 




\subsection{2D Differential Visualization (2-DifViz)}
\label{method3}

To generate realistic and reproducible visualizations of the handwritten digits in air, we developed a method called 2-Dimensional Differential Visualization (\textbf{2-DifViz}).
We use the set of steps below\footnote{This pipeline was built upon the mouse controller application (\url{http://developerblog.myo.com/build-your-own-mouse-control-with-myo/}) developed by Thalmic Labs (\url{https://www.thalmic.com/}).} to get coordinate sequences $C_i$, henceforth referred to as \textit{2-DifViz features}.
These sequences are then plotted on a 2D canvas and are interpolated to smoothen the curve and make the visualizations continuous.

\begin{itemize}
\item \textbf{STEP-1 (Rotate Frame of Reference)}:  $\textbf{g}_i^t$ holds the angular velocity of the user's arm in degrees per second in the three dimensions: $dx$ (\textit{pitch}), $dy$ (\textit{yaw}) and $dz$ (\textit{roll}). These values are in the frame of reference of the arm on which the Myo-armband is worn.
We assume that the digits were written on an imaginary canvas in air, to which we refer as ``the world frame of reference", and hence we rotate $\textbf{g}_i^t$ and bring it to ``the world frame of reference". Let $(\textbf{q}_i^{t})^{-1}$ denote the inverse of $\textbf{q}_i^{t}$. The rotated vector is therefore:
\begin{equation}
\label{eqn:rotateQuat}
\hat{\textbf{g}}_i^{t}=\textbf{q}_i^{t} \star \textbf{g}_i^t
                        \star (\textbf{q}_i^{t})^{-1}
\end{equation}

where $\textbf{g}_i^{t}$ is reinterpreted as a quaternion whose real coordinate is 0, $\hat{\textbf{g}}_i^{t}$ is the rotated gyroscope vector, and $\star$ denotes the Hamilton product.

\item \textbf{STEP-2 (Extract Pitch and Yaw)}: we construct the vector $g_i^{t}=(dx_i^t, dy_i^t)$ from $\hat{\textbf{g}}_i^{t}$, where each $dx$ denotes the \textit{pitch} and each $dy$ denotes the \textit{yaw} in a given time-step.
We ignore the \textit{roll} as we are concerned with mapping the hand movements to a 2-dimensional vector sequence that we can visualize as realistic digits.   
\item \textbf{STEP-3 (Determine the Gain)}: a gain factor $K$ is calculated, that maps the arm movements to pixels on the imaginary canvas. $K$ is determined by hyperparameters like \textit{sensitivity}, \textit{acceleration} and \textit{pixel density} and acts as a scaling factor for $g_i^{t}$.    
\item \textbf{STEP-4 (Calculate Sequence of Differentials)}: $g_i^{t}$ is multiplied with $K$ and a frame duration $F$ to scale the hand movements on a 2-dimensional canvas and smoothen the transitions on it.
\begin{equation}
\label{eqn:difFeat}
\tilde{g}_i^{t}=g_i^{t}\times K\times F
\end{equation}
Here $\tilde{g}_i = \{\tilde{g}_i^1, \tilde{g}_i^2,...,\tilde{g}_i^{\tau_i}\}$ is a sequence of 2-dimensional vectors that contains the number of pixels to move on the imaginary canvas at every time-step. Since $\tilde{g}_i^{t}$ consists of pixels, $dx$ and $dy$ are converted into integers.

\item \textbf{STEP-5 (Create Coordinate Sequences)}: a sequence of 2-dimensional coordinates $C_i^{t}=\{(x_i^1, y_i^1),...,(x_i^{\tau_i}, y_i^{\tau_i}) \}$ is created from $\tilde{g}_i^{t}$, where $x$ and $y$ are coordinates on the horizontal and vertical axis of a 2-dimensional Cartesian plane, respectively, and $|C_i|=\tau_i+1$. To create $C_i^{(t)}$ we start by setting $C_i^{1}=(0, 0)$. Then, $\forall t\in[1,...,\tau_i]$ and $\tilde{g}_i^{t}=(dx_i^{t}, dy_i^{t})$ we set $C_i^{t+1}=(x_i^t+dx_i^t, y_i^t+dy_i^t)$.

\end{itemize}

Figure~\ref{fig:vizHandDigits} shows visualizations of handwritten digits in air using 2-DifViz. This canvas is then stored as an SVG or PNG file. These visualizations were used to create a set $I = \{I_i\mid i=1,...,n\}$ of images, where $n$ is the number of instances in the dataset.

\subsection{GRU Networks}
\label{gruNetwork}

Since Recurrent Neural Networks (RNN) using LSTM have shown state-of-the-art performance for handwriting recognition \cite{bidirLstm, gravesCtc, multiRnn}, we chose to use a similar architecture with a variant of LSTM called Gated Recurrent Units (GRU) for HDRA. RNNs with GRU have fewer parameters than LSTM and their performance is at par with LSTM Networks \cite{GRUequations}. RNNs are able to learn from sequential data and incorporate contextual information, making them a best fit for the HDRA task.

GRUs use gating units as follows \cite{GRUequations}:
\begin{align}
\label{eqn:gru1}
h_t^j &= (1 - z_t^j) h_{t-1}^j + z_t^j \tilde{h}_t^j\\
\label{eqn:gru2}
z_t^j &= \sigma\left( W_z x_t + U_z h_{t-1} \right)^j\\
\label{eqn:gru3}
\tilde{h}_t^j &= \tanh\left( W x_t + r_t\odot\left( U h_{t-1}\right) \right)^j\\
\label{eqn:gru4}
r_t^j &= \sigma\left( W_r x_t + U_r h_{t-1} \right)^j
\end{align}
The activation $h_t^j$ (Equation (\ref{eqn:gru1})) of a GRU at time $t$ is a linear interpolation between the previous activation $h_{t-1}^j$ and the candidate activation 
$\tilde{h}_t^j$, computed by Equation (\ref{eqn:gru3}), where $r_t$ is a set of reset gates and $\odot$ is an element-wise multiplication.
The {\it update gate} $z_t^j$ decides how much the GRU updates its content, according to Equation (\ref{eqn:gru2}).
Finally, the reset gate $r_t^j$ controls how much of the previously computed state to {\it forget} and is computed by Equation (\ref{eqn:gru4}).




The formation of digit in air is dependent on the past as well as the future context and needs to be classified only after the whole digit has been formed. Therefore, we further extended the GRU by combining it with a Bidirectional RNN~\cite{BRNN} resulting in a Bidirectional GRU, known to give better results than  unidirectional RNNs~\cite{BRNNbetter}.
Two GRU networks were trained using the following architectures:

\subsubsection{\textbf{BGRU Network (GRU-NET-1)}}
\label{gruXy}

The 2-DifViz features (\textit{i.e.,}, $C_i$) are post-processed in the following way:

\begin{itemize}
\item \textbf{STEP-1 (Smoothing)}: 
to remove noisy points, each $C_i^{t}$ was smoothed by averaging adjacent points using the following equation \cite{preprocessXY}: 
\begin{equation*}
\hat{C}_i^t=\Bigg(   \frac{x_i^{t-2}+\cdots+x_i^{t+2}}{5},
                     \frac{y_i^{t-2}+\cdots+y_i^{t+2}}{5}    \Bigg)
\end{equation*}
\item \textbf{STEP-2 (Redundancy Removal)}: points too close to each other can convey noise in the direction of movement of a stroke \cite{preprocessXY}. The following equation was used to calculate a threshold $\Delta$, and adjacent points with $\Delta \leq 5$ were removed:
\begin{equation*}
\Delta = \sqrt[]{(x_i^t - x_i^{t-1})^2+(y_i^t - y_i^{t-1})^2}
\end{equation*}
\item \textbf{STEP-3 (Standard Scaling)}: $C_i^{t}$ is whitened by scaling its values by its mean and standard deviation.
\begin{equation}
\label{standardScaling}
 \hat{C}_i^{t}=\frac{C_i^{t}-\mu_{c}}{\sigma_{c}}
\end{equation}

Where $\mu_{c}$ and $\sigma_{c}$ are the mean and standard deviation of $C_i$, respectively. This helps in removing the offset of the sequences by making $\mu=0$ and normalizes the fluctuation by making $\sigma=1$.
\item \textbf{STEP-4 (Interpolation)}: all coordinate sequences $C_i$ are linearly interpolated to the same length $|\hat{C}_i|=100$
\end{itemize}

The newly generated $\hat{C}_i$ are used to train a 1-layer BGRU network with 32 output units with a sigmoid activation function. A softmax output layer was used with 10 units for the 10 digits. We trained the network using Stochastic Gradient Descent (SGD) over the training data for 150 epochs with a categorical crossentropy loss and the Adam optimizer with a learning rate of 0.001, $\beta_1=0.9$, $\beta_2=0.999$, $\epsilon=10^{-8}$ and decay of $10^{-6}$.

\subsubsection{\textbf{BGRU Network (GRU-NET-2)}}
\label{gruImu}

All the raw $\mathcal{M}_i$ are standardized in the following way:
\begin{itemize}
\item \textbf{STEP-1 (Absolute Scaling)}: all $\mathcal{M}_i^t$ are scaled to a range of $[-1,1]$.

\item \textbf{STEP-2 (Resampling)}: a set $T=\{\tau_i\mid1,...,n\}$ is defined such that $n=|\mathcal{D}|$ and $\tau_i$ is the number of time-steps in $\mathcal{M}_i$. Let $t_{max}$ be the maximum of all values in $T$. For each $d_i$, the time-series $\mathcal{M}_i$ is resampled such that  $|\mathcal{M}_i|=t_{max}$. This is done to normalize the length of the sequences and remove jitter.
\end{itemize}

We trained a 1-layer BGRU on the standardized $\mathcal{M}_i$. The number of units, activation, output layer, loss and optimizer used were the same as the GRU-NET-1.

\subsection{Convolutional Neural Network}
\label{cnn}
Convolutional Neural Networks (CNN) have shown state-of-the-art results in classifying images\cite{mnistCnn,cifarCnn,mnistCnn2}. Since we were able to convert the handwritten digits in air into realistic visualizations (images), we could reduce the HDRA problem to an image recognition problem. CNNs have outperformed other existing models for the MNIST digit recognition task \cite{mnistCnn} so we chose to use CNNs for HDRA using 2-DifVis. We used a transfer learning approach \cite{surveytransferlearning} for re-training a pre-trained CaffeNet\footnote{\href{https://github.com/BVLC/caffe/tree/master/models/bvlc-reference-caffenet}{https://github.com/BVLC/caffe/tree/master/models/bvlc-reference-caffenet}} \cite{caffee} on the image set $I$. The output layer of the CaffeNet was replaced with a softmax layer with 10 units and the network was re-trained using the Adam optimizer with a learning rate of $0.0001$, $\beta_1 = 0.9$, $\beta_2 = 0.999$ and $\epsilon = 10^{-8}$ using a categorical cross entropy loss. 
\subsection{Fusion Model}
By using GRU-NET-1 and GRU-NET-2 we were able to capture the temporal information from the preprocessed coordinate sequences and IMU data. With the CNN we could acquire a spatial representation of images generated by 2-DifViz. This motivated us to extend our model and fuse these three modalities to capture the spatio-temporal representation of the raw IMU data. To do so, we fused the ranked results using Borda Count, whereby we extracted the ranked class labels from the three classifiers and decided a final class label for the input. Since the modalities are independent of each other, the sources of errors are independent too. This makes our fusion model robust and a best fit for the HDRA task.

%
 

\section{Evaluation}
\label{evaluation}
We evaluated our method using a person dependent test and a person independent test, as described in the sections below. To benchmark the components of the fusion model we compared their accuracies with a Hidden Markov Model (HMM), a Support Vector Machine (SVM), a Naive Bayes (NB) classifier and a K-Nearest Neighbor (KNN) classifier. 

\subsection{Person Dependent Test}
To perform a person dependent test, the data from a single participant is used to train and evaluate the model. We split the person's data into 5 stratified folds. Each fold consists of a training set and a test set. We repeat this process for 10 randomly selected participants. We train and evaluate all our models on all the folds from each of the selected participants. The average accuracy on the 5 folds for each participant is recorded and a mean of these averages is calculated, as well as their standard deviation.  Table \ref{tab:pdependent} shows the results for all the evaluated models.


\begin{table}
\caption{Mean average accuracy and standard deviation of the person dependent evaluation.
\label{tab:pdependent}
}
\begin{center}
\begin{tabular}{l|c|c}
Classifier & Avg. Mean Accuracy (\%) & Std. deviation
\\
\hline
 	HMM					&	 75 	 	&	 	22.9 	\\
 	KNN     			&	 77.5	 	& 		16.9	\\
	NB      			& 	 75.3	 	&		21.3	\\
	SVM     			& 	 81.67		&		13.9	\\
    CNN 				& 	 95.1 		& 		5.9		\\
	GRU-NET-1 			& 	 84.4 		& 		13.4	\\
	GRU-NET-2 			& 	 88.7 		& 		10.6	\\
	\textbf{Fusion Model} 		& \textbf{96.7} &       \textbf{0.02} \\
\end{tabular}
\end{center}
\end{table}


\subsection{Person Independent Test}
We use this test to evaluate the robustness of a classifier, independently of any specific person. In this setting, we withhold the data of a randomly selected participant to be used as a test set, while the rest of the data is used as a training set. This is repeated for 10 different participants and the average accuracy is recorded. Table \ref{tab:pindependent} shows the average accuracy of all the evaluated models. 

\begin{table}
\caption{Mean average accuracy and standard deviation of the person independent evaluation.
\label{tab:pindependent}
}
\begin{center}
\begin{tabular}{l|c|c}
Classifier & Mean Accuracy (\%) & St. Deviation\\
\hline
 HMM
                                         & 15.8 & 5.8 \\
 KNN            
                                         & 13.6 & 8.8  \\
NB       & 31.7 & 13.6  \\
SVM    & 19.7 & 13.8 \\
CNN  & 84.6 & 11.2\\
GRU-NET-1 & 67.4 & 10\\
GRU-NET-2 & 87.6 & 10.4 \\
\textbf{Fusion Model} & \textbf{91.7} & \textbf{0.06}\\
\end{tabular}
\end{center}
\end{table}


\begin{figure}[!t]
\includegraphics[width=.5\textwidth]{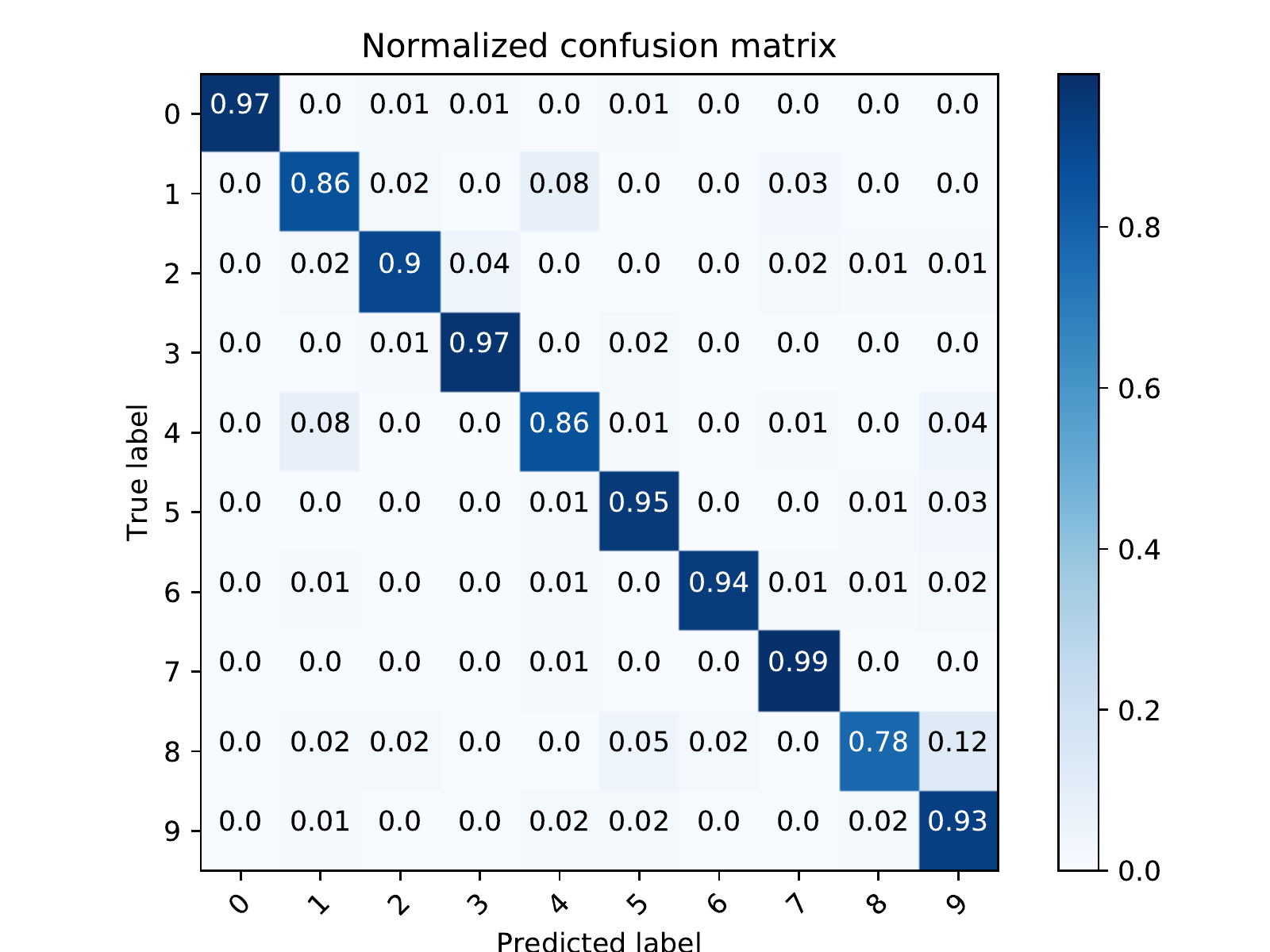}
\caption{Confusion matrix for HDRA using the fusion model.}
\label{fig:confusion}
\end{figure}


\subsection{Analysis}
\subsubsection{Performance Analysis}
In both the person dependent and person independent tests, the fusion model outperformed all the baseline models with a substantial margin. The CNN, GRU-NET-1 and GRU-NET-2 were the top 3 classifiers considering the average accuracy. Thus, we chose to use them in our fusion model, which resulted in an improvement in accuracy. Even though the CNN, GRU-NET-1 and GRU-NET-2 had an accuracy of more than $80\%$, they had a standard deviation of more than 10. However, fusing them together resulted in a drop in standard deviation, making the fusion model much more robust.

We observed that the results were better in the person dependent test than the person independent test. This behavior can be attributed to each person having different speed and style of writing the same digits. 
\subsubsection{Digit Analysis}
We analyzed the confusion matrices of the fusion model and its three components individually. For doing so, we relied on the 2-DifViz visualizations of the hand movements of the participants. This analysis helped us in improving the classifiers and observing the subtle problems in digit recognition. The confusion matrix of the fusion model is shown in Fig. \ref{fig:confusion}.
The confusion matrices of the different classifiers allowed us to have an insight on what errors they were committing.
For instance, digits $1$ and $2$, as well as $1$ and $4$ were confused in GRU-NET-2, possibly because of similar initial hand motion. Similarly, the CNN tended to confuse digits with loops (like $6$, $8$ and $9$). Fusing the models together allowed us to overcome these difficulties and achieve a higher robustness. 

\section{Conclusion and Future Work}
\label{conclusion}
We introduce a novel approach called AirScript for creating documents in air using a Myo-armband. For doing so, we split the problem into a visualization task (2-DifViz) and a handwriting recognition task (HDRA). We show a proof of concept for the latter by proposing a classifier fusion model which achieves a recognition rate of $91.7\%$ on a person independent evaluation, and $96.7\%$ on a person dependent evaluation. For the visualization task we introduce a new method called 2-DifViz, which converts the hand movements into realistic visualizations on a 2D canvas of the digits written in air, that can be stored in an SVG or PNG format. This shows the potential use of AirScript in many application areas such as, smart factories, smart offices, smart class rooms, virtual reality games and even augmented reality environments. We envision AirScript to be used in a smart classroom environment using augmented reality, where people can scribble anything as air notes and visualize these notes in the form of handwriting, thus giving the process of creating a new definition. AirScript uses a Myo-armband, our method is mobile and easy to integrate on multiple platforms, while still providing the user with freedom of movement.

We plan to extend AirScript by adding a handwriting recognition model using Sequence to Sequence models or LSTM Networks with Connectionist Temporal Classifiers, along with a language model. We plan to use Deep Generative Models for improving the visualizations of handwriting in air and making them more realistic.


\section*{Acknowledgment}
We would like to thank Software-Labor, TU Kaiserslautern for providing us with a Myo-armband and a working space. We would also like to thank MindGarage, TU Kaiserslautern for providing computational resources.



\bibliographystyle{IEEEtran}
\bibliography{IEEEabrv,main}
%

\end{document}